\documentclass{article}

\usepackage{arxiv}

\usepackage[utf8]{inputenc} 
\usepackage[T1]{fontenc}    
\usepackage{hyperref}       
\usepackage{url}            
\usepackage{booktabs}       
\usepackage{amsfonts}       
\usepackage{nicefrac}       
\usepackage{microtype}      
\usepackage{lipsum}		
\usepackage{graphicx}
\usepackage{natbib}
\usepackage{doi}

\title{Systematic Adaptation of Communication-focused Machine Learning Models from Real to Virtual Environments for Human-Robot Collaboration}

\author{ \href{https://orcid.org/0000-0003-4950-0631}{\includegraphics[scale=0.06]{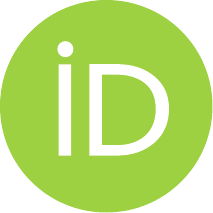}\hspace{1mm}Debasmita Mukherjee}\\
	School of Engineering\\
	The University of British Columbia\\
	Kelowna, BC V1V 1V8 \\
	\texttt{debasmita.mukherjee@alumni.ubc.ca} \\
	\And
	\href{https://orcid.org/0000-0000-0000-0000}{\includegraphics[scale=0.06]{orcid.pdf}\hspace{1mm}Ritwik Singhai} \\
	Department of Aerospace Engineering\\
	Indian Institute of Technology, Kharagpur\\
	Kharagpur, West Bengal 721302 \\
	\texttt{ritwik.singhai@iitkgp.ac.in} \\ 
	\And
	\href{https://orcid.org/0000-0002-3550-225X}{\includegraphics[scale=0.06]{orcid.pdf}\hspace{1mm}Homayoun Najjaran}\thanks{Corresponding author} \\
	Department of Mechanical Engineering\\
	University of Victoria\\
	Victoria, BC V8P 5C2 \\
	\texttt{najjaran@uvic.ca} \\
}

\renewcommand{\shorttitle}{\textit{Systematic Adaptation of Communication-focused ML from Real to Virtual for HRC}}

\hypersetup{
pdftitle={A template for the arxiv style},
pdfsubject={q-bio.NC, q-bio.QM},
pdfauthor={David S.~Hippocampus, Elias D.~Striatum},
pdfkeywords={First keyword, Second keyword, More},
}

\begin{document}
\maketitle

\begin{abstract}
Virtual reality has proved to be useful in applications in several fields ranging from gaming, medicine, and training to development of interfaces that enable human-robot collaboration. It empowers designers to explore applications outside of the constraints posed by the real world environment and develop innovative solutions and experiences. Hand gestures recognition which has been a topic of much research and subsequent commercialization in the real world has been possible because of the creation of large, labelled datasets. In order to utilize the power of natural and intuitive hand gestures in the virtual domain for enabling embodied teleoperation of collaborative robots, similarly large datasets must be created so as to keep the working interface easy to learn and flexible enough to add more gestures. Depending on the application, this may be computationally or economically prohibitive. Thus, the adaptation of trained deep learning models that perform well in the real environment to the virtual may be a solution to this challenge. This paper presents a systematic framework for the real to virtual adaptation using limited size of virtual dataset along with guidelines for creating a curated dataset. Finally, while hand gestures have been considered as the communication mode, the guidelines and recommendations presented are generic. These are applicable to other modes such as body poses and facial expressions which have large datasets available in the real domain which must be adapted to the virtual one.
\end{abstract}

\section{Introduction}
Human-robot collaboration (HRC) studies have increasingly found that the presence of a robot in the room exerts a similar effect on humans as having another human in the room (i.e., social facilitation) in terms of task performance and elicitation of feelings of a “social presence” during the interaction \cite{i1}. Social presence is defined as the feeling of being aware of another social being in the context of human-computer or human-machine communication. Indeed, co-present or “embodied” robots have been determined (for certain interaction tasks) to assert peer pressure on humans more effectively than virtual agents, especially for tasks with high uncertainty \cite{i2}. The physical embodiments of robotic agents are available in a wide range of levels of anthropomorphism which significantly affects trust between humans and robots \cite{i3}. If the motivation of HRC systems is to encourage effective cognitive decision-making from human participants, ease of interaction and engagement in joint tasks are important considerations. Thus, embodiment of the robot partners in the task is an essential design requirement. This assumption must hold true for teleoperation scenarios wherein the interaction or collaboration between humans and robots occurs between agents separated by geographical distance. In order to engage the humans more effectively in such cases, embodiments of both agents in a virtual reality environment may be effective.  

Virtual reality (VR), once a niche technology, has enabled the development of a plethora of interactive applications in a variety of settings, engaging postures \cite{engPostures}, gestures \cite{gandspeech}, gaze \cite{gaze}, and speech \cite{gandspeech} across diverse applications \cite{vrgen}. Besides its use in gaming, VR is increasingly being used in HRC applications to provide multimodal inputs to the robot \cite{hjorth2022human}. This would enable the robot to carry out non-ergonomic tasks allowing the human partner to focus on tasks requiring precision and cognitive decision-making \cite{MUKHERJEE2022102231}. Along with the traditional tasks, VR and the flexibility of designing a virtual environment empower designers to explore applications outside of the constraints posed by the real world environment and develop creative designs, innovative solutions and immersive experiences. 

For successful deployment of HRC systems in industries, especially to engage workers who are not specially trained, the mode of communication must be easy to learn and as close to “natural” as possible \cite{Mukherjee3, Mukherjee_DST}. The usage of a VR environment and development of virtual agents will thus meet the two-fold requirements of providing copresence of the robotic agent as well as a “natural” manner of interaction. That said, while integration with simulation and development platforms such as Unity have enabled the rapid creation and editing of immersive environments, programming of new postures and gestures for rapidly evolving applications is still a time-consuming task. The inclusion of more and more gestures or postures through the means of combinations of buttons on the VR controllers would thus become increasingly more complicated and may distract the workers from engaging their optimal cognitive faculties towards task decision-making. One means of alleviating this likely problem would be the inclusion of computer vision through data-driven techniques such as machine learning (ML) \cite{chai2021deep} to interpret data generated in the VR environment and captured through virtual “cameras”.

Over the past decade, sophisticated ML architectures have been developed to leverage features acquired from large datasets to carry out tasks such as detection \cite{dhillon2020convolutional} and recognition \cite{he2016deep}. Immense resources have been dedicated to creating labeled datasets for communication modes, such as hand gestures, human poses, voice commands, and facial expressions \cite{MUKHERJEE2022102231}, that enable models to achieve high accuracy values in recognition tasks. These ML models facilitate the conveyance of information to the collaborating robot. Apart from that, the dynamic nature of a workcell or work place setting necessitates flexibility of recognition of commands or information tackling a variety of noise. Well-trained ML models developed using large, labelled datasets utilize their ability to generalize (to an extent) and approximate to understand such variable scenarios.
Leveraging such architectures for the purpose of communication in the VR realm faces the stumbling block of data scarcity. For an HRC, or more generally, a human-machine interaction scenario, a VR environment (VEnv) linked to the real-world environment (REnv) may contain a representation of the human in it to facilitate copresence with the robot in a teleoperation task. The first step to that end is to establish a perception system that would allow communication between agents. 

In the present work we have considered hand gestures as the mode of communication. The real gestures in REnv may be re-created in the VEnv which would then be recognized using an ML model, much like REnv gestures. Thus, the VEnv may be conceived as composed of two parts – the representation of the human through a \textit{humanoid} model and the setting of the immersive \textit{scene} which includes the light(s), the virtual sensor(s), other interacting object(s) and artifacts such as walls, floor, ceiling, their colours and so on. For developers, the construction of the \textit{scene} is arduous but that of the \textit{humanoid} with the complicated gestures is harder still, consuming large resources and time. Gestures which seem simple and ubiquitous in REnv need to be meticulously created by programming multiple joints. In order to simulate the dynamic nature of the industrial setting or workcell, variability in the positioning of the\textit{ humanoid} as well as variability if the positioning of the hand, the degree to which the gestures are performed and the setting of the workcell itself must be represented in the dataset. This proves to be a challenging task and must be managed in order to leverage the power of ML methods.

\section{Related Works}
This scarcity of datasets in one domain may be alleviated by bringing pertinent knowledge from another, more mature domain through what is called Domain Adaptation. This is used to the bridge the gap between the source and the target domains. Several works have been undertaken to adapt datasets generated in the virtual domain to the real. Classifiers based on neural networks are discriminative models that are trained on appearance as core feature. The work presented in \cite{rw1} and \cite{rw2} aimed to use a pedestrian detection model trained using images generated from "realistic virtual scenarios" from players of a video game to detect the same in real-world settings. Even if the virtual world-generated images are photo-realistic, there is a marked dataset shift due to the difference in behaviors of the virtual and real cameras. Therefore, researchers in the cited work proposed a domain adaptation framework, V-AYLA, which combined images from the real and virtual world and showed similar performance to model trained only with real-world images. Similarly, in \cite{rw3}, 2D object detection was carried out using model trained with3D models rendered using virtual data. A virtual-to-real scene parsing was trained using a novel few-shot structured domain adaptation model with small number of target real images and associated semantic labels \cite{rw4}. These were collaborated with data from the virtual domain. The model contained a two-stage adversarial network with a scene parser and two discriminators. 

In the VR-enabled embodied teleoperation scenario envisioned here is the reverse of the afore-mentioned cases. Large, annotated hand gesture datasets (in REnv) have been extensively created and used to train models achieving high accuracy values. The goal is to undertake domain adaptation of a model trained using such a standard dataset such that it is operational in the VEnv.

\section{Contributions}
The usual practice in designing ML models for computer vision is to select the ML model architecture and training parameters according to the dataset aided by decades of research invested. In this case however, the adaptation from real to virtual is the reverse scenario. As mentioned earlier, designers have control over all aspects of the virtual environment (VEnv) through manipulation of such parameters as the number of polygons, the details of lights, textures, the physics associated with objects and animation \cite{vrgen}. The limitation to achieving photorealism is time and computational resources. In our application, the humanoid and gestures are more challenging to design than the environment. This poses the constraint of developing a framework that can function on small datasets of VEnv data that must be curated to work with the well-performing ML model that has been trained on the communication mode that is to be used. This would allow us to leverage the capability of classification of complex gestures and further to personalize the ML models generated to the interacting human as presented in \cite{Mukherjee3, thesis}.

The paper thus presents guidelines to adapt a well-performing model in the REnv to perform well in the virtual domain without the use of a large dataset through the real to virtual adaptation (R2VA) framework through a systematic curation of VEnv datasets. The contributions constitute an isagoge to designing of curated datasets in the virtual environment that is representative of the real-world communication and can be used with a machine learning model trained on the real data achieved through the use of interpretable artificial intelligence techniques and iteratively editing the dataset and testing with the model. The presented work focuses on the adaptation of the dataset while future works will delve into linking the real-world hand gestures to the virtual. The results in this paper have been obtained with the iterative editing of RGB images of the\textit{ humanoid} and the \textit{scene}. Finally, while hand gestures have been considered as the communication mode, the guidelines and recommendations are generic and are applicable to other modes such as postures and facial expressions that can be recognized through images or videos and have large datasets available in the real domain with the creation of similar sized ones in the virtual domain being resource heavy.

\section{Real Environment to Virtual Adaptation (R2VA) Framework}
The real environment (REnv) to virtual (VEnv) adaptation framework (Figure 1) for communication modes would first require a well-performing model (\textit{model0}) trained with a REnv dataset. Such a model must possess large variance and small bias. A VEnv dataset is created \textit{ad hoc} using the available 3D models or assets in the environment and with an initial structure of the scene. This dataset is used as a test set on \textit{model0}. If the classification performance, measured through accuracy is acceptable, the similarity in features (in terms of components of the VEnv i.e., the \textit{humanoid} and the \textit{scene} are noted. The model may be deployed into the VEnv for communication applications. On the other hand, if the performance is below the set threshold, interpretable artificial intelligence (AI) tools may be used to better understand the results. These would allow interpretation of the features learned by \textit{model0} and that it expects to find in the test set so as to correctly classify the data. These inferences form guidelines for further design of the \textit{scene} and/or the \textit{humanoid}. A checkpoint of essence that needs to be added to any such framework is whether it is feasible in terms of time, skill, and computational resources to carry out the changes to the dataset. If it is indeed feasible, then the features may be added and the VEnv dataset tweaked to better resemble the trained dataset. The entire process of testing is carried out again iteratively and the dataset improved upon till it becomes inviable to make any more changes. At this juncture, transfer learning may be carried out on \textit{model0} using the latest dataset created. Following the work presented in \cite{Mukherjee4, thesis}, datasets with small number of images are enough to compensate for the real to virtual domain shift. Finally, the new model generated, \textit{model1} may be deployed to the VEnv with the VR environment. While rule-based systems or more traditional forms of AI could be used without following the presented framework, the versatility of VR and the flexibility it grants in changing all aspects of the immersive environment requires an ML model that contains enough trainable parameters to learn features in future changes of the sensors, the structure of the \textit{scenes}, the \textit{humanoid}, and the communication mode. This requirement along with the constraint posed by small datasets provides justification for the use of the R2VA framework. The following section presents an application of the same to the task of hand gesture recognition for an HRC scenario.

\begin{figure}
      \centering
      \includegraphics[scale=0.6]{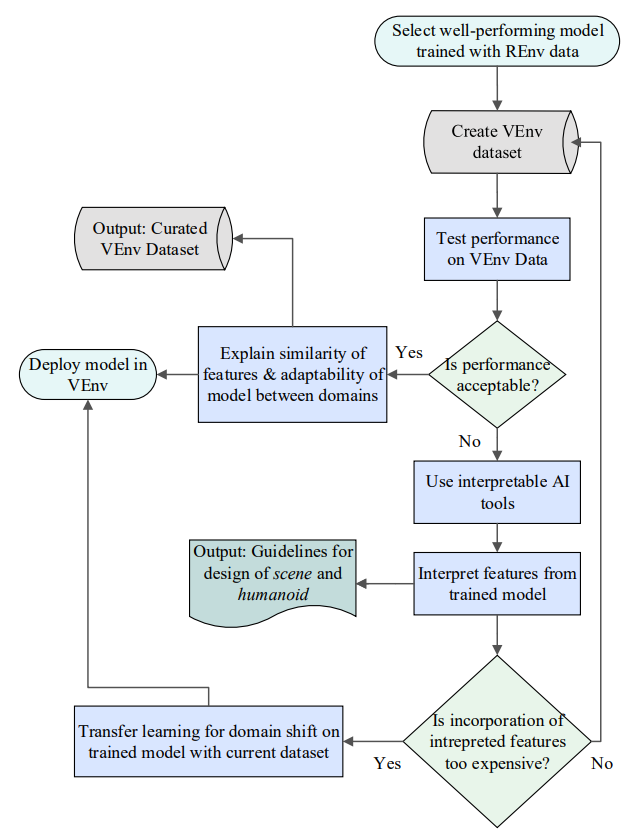}
      \caption{Real Environment to Virtual adaptation (R2VA) Framework}
      \label{figurelabel}
\end{figure}

\section{Case Study for Application of R2VA Framework}
Hand gestures were selected as the mode of communication since they are a natural form of communication. A number of gestures are ubiquitous using which do not require providing users or workers in industries with any specialized training. Several large hand gestures datasets are publicly available and extensive research has been carried out in hand gesture recognition \cite{ MUKHERJEE2022102231}. 

\subsection{Selection of well-performing Model trained with REnv Data}
For the presented work, VGG 19 \cite{vgg} was trained with a publicly available hand gesture dataset: Image database for tiny hand gesture recognition \cite{TH}, hereon referred to as TH. The training parameters used were 0.01 learning rate, 16 batch size, and stochastic gradient descent optimizer run till 20 epochs. Data from four of the participants was used as test set while the test: validation split was (70:30) \%. The dataset used is composed of RGB images with seven hand gestures acted out by 40 participants. These hand gestures may be considered quite typical for communicating with a robot or other intelligent system. The classes of hand gestures are: {\textit{fist, l, ok, palm, pointer, thumb down, and thumb up}}. Variability has been added by the creators of the dataset by incorporating gestures performed in different locations in the images as well as by having the participants positioned with a number of different backgrounds both cluttered as well as clean, solid coloured, and with variation in illumination. The human face and body occupy the majority of the image while the hand occupies only about 10\% of the pixels of the image. In a dynamic industrial environment, the human may not always be positioned in a central location or have the hands close to the camera, thus, the dataset was selected to mimic that inconsistency in sensor data.

The trained VGG19 model (\textit{model0}) was then tested on the TH test set and had an accuracy of 0.75. In order to test the model on the same hand gestures carried out by a human originally not a part of the original dataset, a user dataset (UDS) was created with a cluttered background similar to a portion of the images in TH. An \textit{ad hoc} VEnv dataset (VDS1) was also created using a silver-coloured \textit{humanoid} carrying out all the gestures in the above-mentioned classes except \textit{pointer }since it was determined to not be a very intuitive way of “pointing”. The accuracy values on UDS and VDS1 were 0.2 and 0.09 respectively. Since these were quite low, SHAP values as an interpretable-AI tool was used to garner more information about the classification results and the behaviour of \textit{model0}. 

SHAP, which stands for Shapely Additive Explanations, was proposed by Scott and Su \cite{shap1} as a unified measure of feature importance. They are the shapely values of a conditional expectation function of the original model. The three core equations from cooperative game theory—namely, Shapley regression values \cite{shap2}, Shapley sampling values \cite{shap3}, and Quantitative Input Influence \cite{shap4} are used by the model interpretability techniques to derive explanations of model predictions. The calculation of shapely values becomes more complex as more features are included, thus requiring the need for approximation methods. A faster Deep Neural Network specific SHAP values approximation method Deep SHAP \cite{shap1} was used in this work which was devised by connecting shapely values and DeepLIFT. In order to calculate SHAP values for the whole network, the Deep SHAP approach combines SHAP values obtained for smaller network components. The DeepLIFT multipliers are iteratively propagated backward down the network and expressed in terms of SHAP values.

SHAP values were obtained for the test images of TH, UDS and VDS1 to get a better empirical understanding of what\textit{ model0} expected to find in the image that helped or hindered in classification. The images from the datasets are presented in the figures with higher colour saturation from the original passed to the ML model so as to make the SHAP regions more legible. Figure 2. (a) presents indicates that \textit{model0} had overfitted to TH data and expected to find the human in the centre of the \textit{scene} as evidenced by the presence of pink regions. This led to wrong classifications on VDS1. \textit{model0} may also have looked for human ‘skin tones’ as a feature since the parts highlighted in pink are the similar coloured background of (a) and (b) in Figure 2. In TH test set, examples of which are in Figure 2. (c) and (d), the model expected to find the hand gesture with respect to the rest of the human body, predominantly the face and the arm, hence these structures are seemingly important parts of the scene. Figure 2. (e) is from the user dataset with a cluttered background and here too the model sought the aforementioned structure. UDS also contained the same gestures carried out with the left hand instead of just the right hand as was done in TH. The accuracy for the left-handed gestures carried out by the user was 0.21. This was to be expected since the model had not been trained for these variations. The background was not treated as a feature, merely as noise which is a desirable behaviour of the model. Notwithstanding the generalizability across different backgrounds, it was concluded that \textit{model0} possessed low variance and low domain adaptability.

\begin{figure}[] 
      \includegraphics[scale= 1]{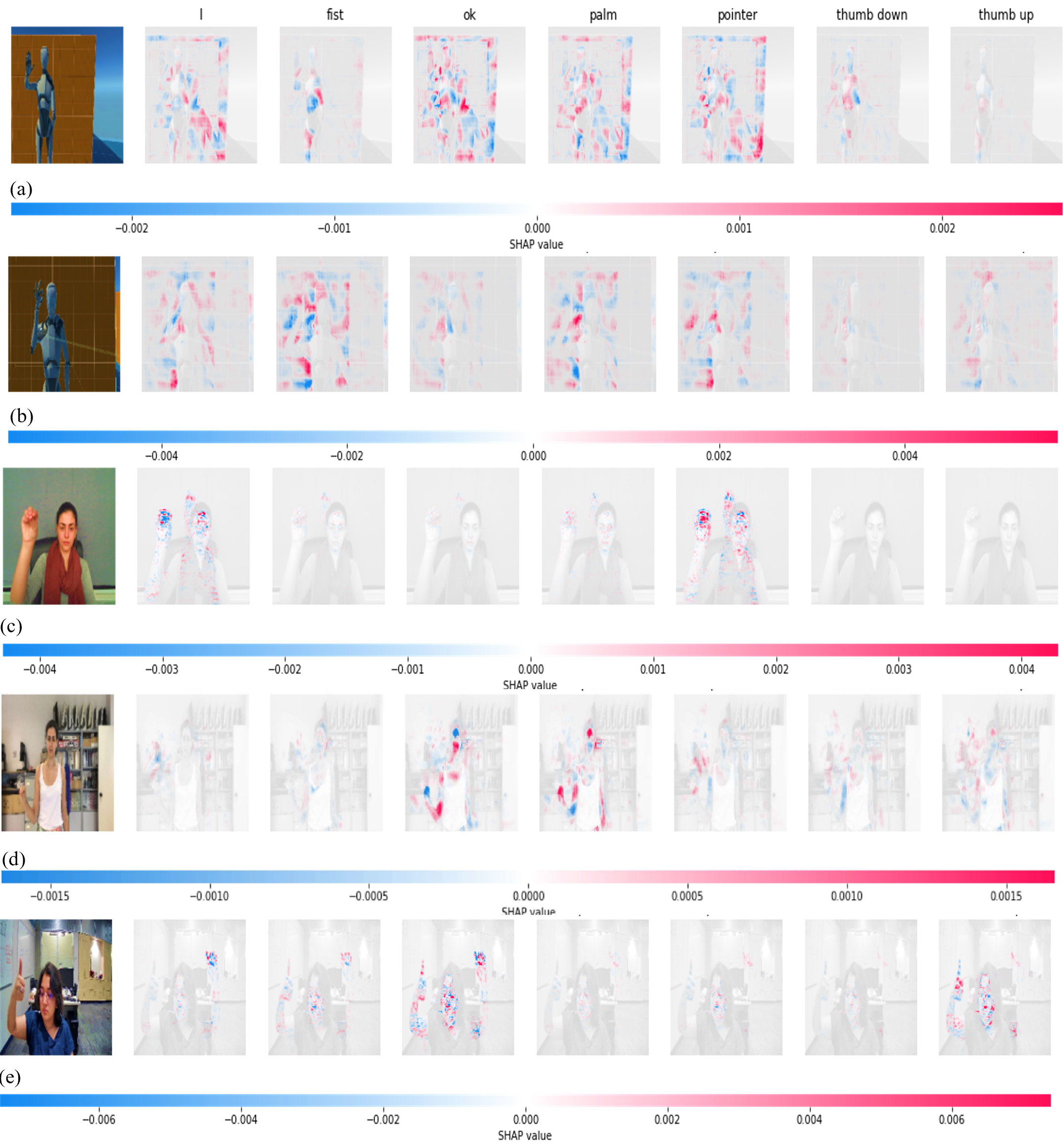}
      \caption{SHAP values of test images of TinyHand (TH), User Dataset (UDS) and first iteration of Virtual hand gestures Dataset (VDS1) tested on \textit{model0} indicating overfitting to: (a) human in centre of \textit{scene} in VDS1, (b) skin tones in VDS1, (c) and (d) hand gestures' relative position to the human body in TH, (e) hand gestures' relative position in UDS}
      \label{figurelabel}
   \end{figure}

A new VGG19 model was trained with extensive data augmentations carried out on TH training set. These included – lateral inversion to include left-handed gestures, lateral shifts so as to not overfit the model to expecting the human to be at the centre of the scene, shears so that the fish-eye effect of certain cameras could be trained for, zoom to mimic the varying distance of the human from the camera in a dynamic environment, small rotation and brightness shifts to allow for dynamic lighting conditions. The test accuracies on TH, UDS and VDS1 were 0.82, 0.59, and 0.08 respectively which were improvements for TH and UDS while not as much for VDS1. The new model,\textit{ model0\_1} was tested with certain variations of TH test images. Figure 3. (a) shows that \textit{model0\_1} still expected the structure and relative positioning of the hand and the human. Without skin tone as a feature in (b), this behaviour is quite evident. In the negative red channel image in (c) and negative green in (d), the parts of the images that contained skin tone like colours were highlighted along with the hand and the arm and body. In (e), without any skin tones, very few portions of the image were highlighted which subsequently led to incorrect classification.

   \begin{figure}[] 
      \includegraphics[scale=1]{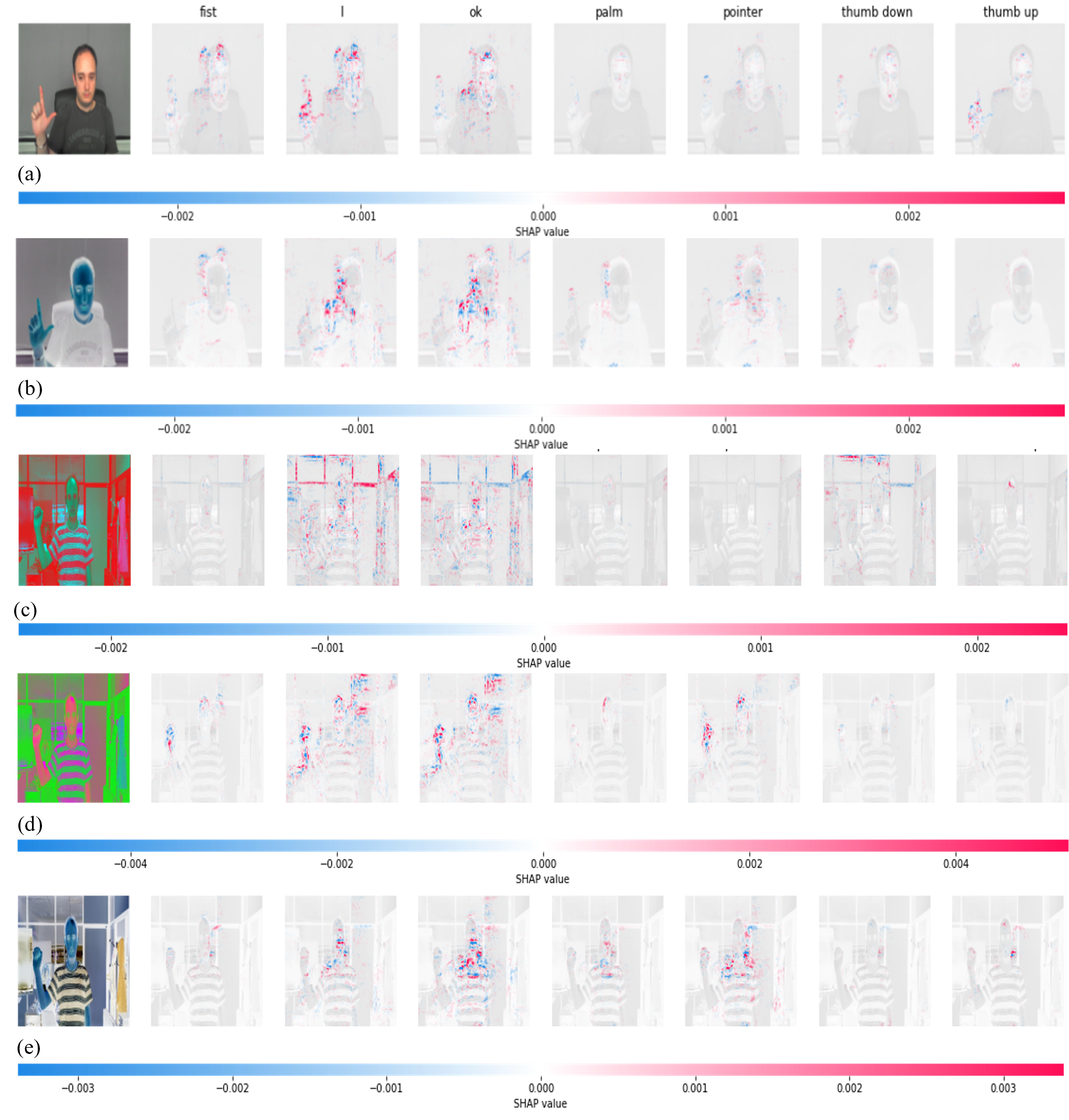}
      \caption{SHAP values of test images of TinyHand (TH) tested on \textit{model0\_1} indicating: (a) and (b) model expecting relative position of hand with respect to the body of the human, (c) and (d) "skin tone" like portions of the images highlighted indicating overfitting to that colour feature, (e) lack of skin tone like colour may have led to incorrect classification}
      \label{figurelabel}
   \end{figure}
   
On investigating more into the behaviour of structure expected by the model, Figure 4. (a) shows that the monitor in the background along with the skin tone like colour of the wall led to the correct classification of \textit{l} but not of the gesture itself but with the monitor. This is because most images in the TH training set contained the hand positioned in the upper left of the image. If the image was cropped to exclude the monitor, the model looked for the relative position of the hand with respect to the human body (Figure 4. (b)). Finally, by obscuring the monitor without cropping the image, the correct prediction of \textit{l} was obtained when the model used the relative position of the hand with respect to the human in (c). Thus, it may be concluded that \textit{model0\_1} expected to find either the gesture and the skin tone or the structure of the \textit{scene} which contained the gesture relative to the human’s head and arm. This behaviour was retained in the model and not treated as an overfitting anomaly since it was surmised that it would be advantageous in an HRC operation wherein the robot would be required to classify the hand gestures of the “main” collaborating human in its field of view and ignore other hands if they happened to be in the scene. Finally, Figure 4. (d) indicates recognition of left-handed gestures, (e) and (f) that it is not overfitted to have the human in the centre of the scene, and (g) that the model gets confused by skin tones. This model was determined to be well-performing and thus used for further steps.

\begin{figure}[] 
 \includegraphics[scale=1]{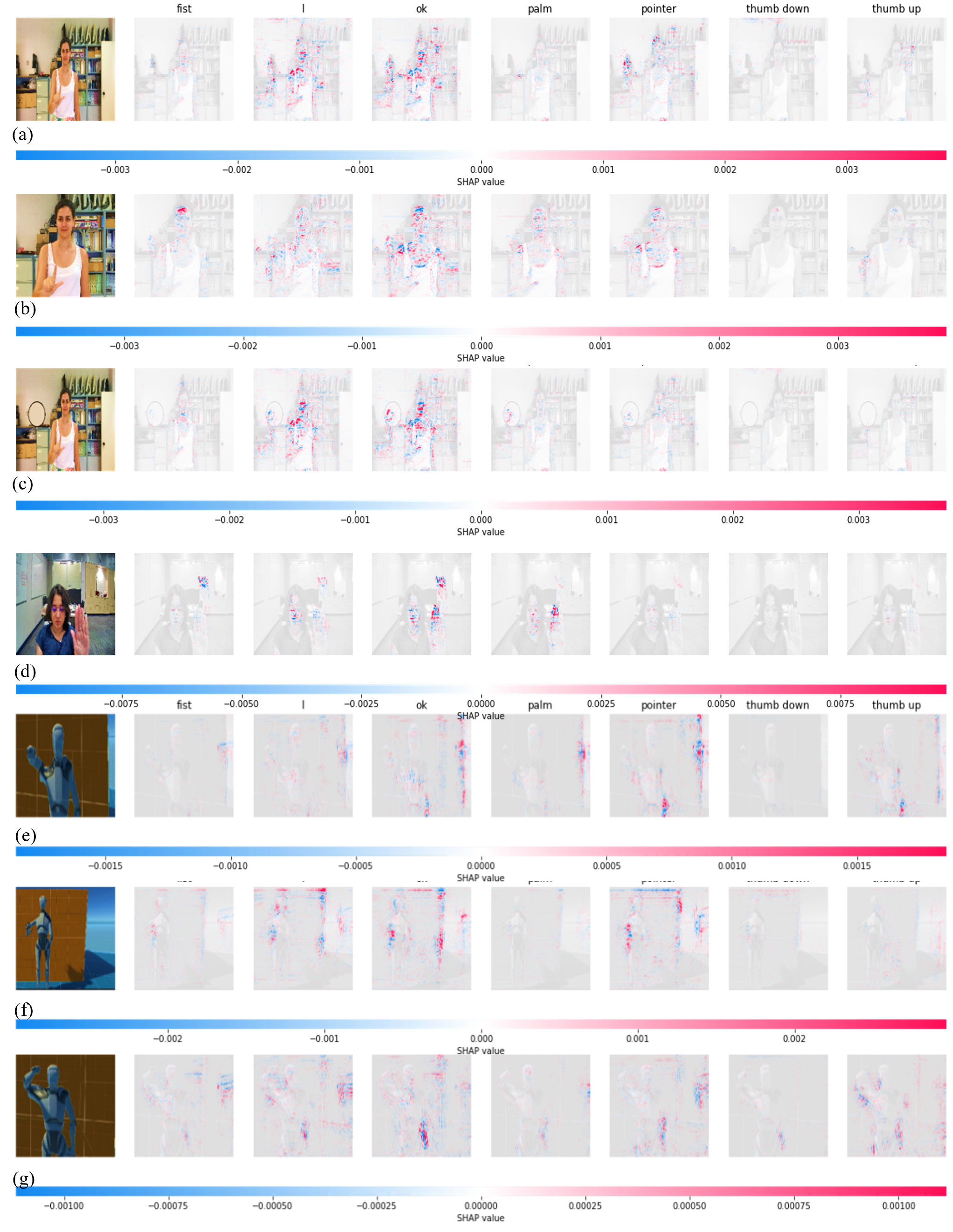}
  \caption{SHAP values of test images of TinyHand (TH), User Dataset (UDS) and first Virtual hand gestures Dataset (VDS1) tested on \textit{model0\_1} indicating: (a) correct classification utilising skin tone and structure of gesture in background element instead of the human hand in TH, (b) use of relative position of hand in cropped image, (c) correct classification with monitor obscured, (d) correct classification of left handed gesture in UDS, (e) and (f) variance in position of figure in images from VDS1, (g) model still using skin tone as a feature}
\label{figurelabel}
\end{figure}

\subsection{Creation of VEnv Dataset }
The inferences from the SHAP values obtained were that the \textit{humanoid} could be placed in varying positions in front of the camera with varying backgrounds, while the colour would have to be changed from silver colour to one more resembling the human skin tone. For best results, the colour of the background should not resemble skin tones. Incorporating these to create a new VEnv dataset (VDS2) yielded significantly better results. The SHAP values indicated that the \textit{humanoid} was recognized as the “human” figure carrying out gestures as can be observed from Figure 5 (a)-(c). Figure 5 (d) contains the \textit{humanoid’s} arm without the body and the SHAP values indicate the recognition of the arm. The accuracy on testing with \textit{model0\_1} was 0.17 which was an increase from the previous test on \textit{model0} but still required improvement. On obscuring both the \textit{humanoid} as well as the human, the SHAP values indicated that the model expected the body of the human which is captured in Figure 5 (e) and (f), thus the VEnv dataset would not work if it merely contained the arms of the humanoid.
 
\begin{figure}
      \includegraphics[scale=1]{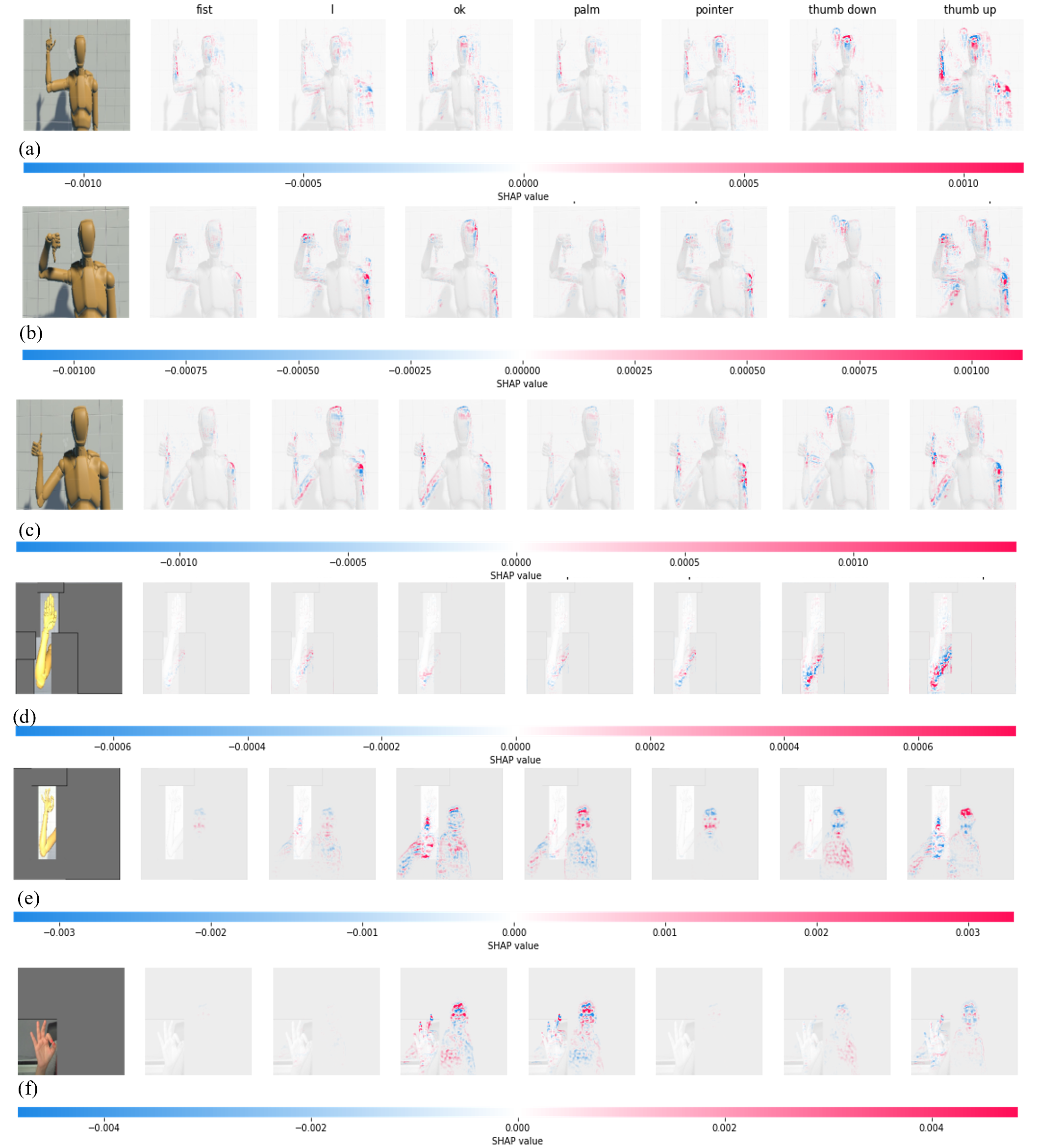}
      \caption{SHAP values indicating: (a)-(c) that the \textit{humanoid} is recognized as the human figure in VDS2, (d) and (e) recognition of arm of the \textit{humanoid}, (e) and (f) the model's expectation of the rest of the human body as a feature}
      \label{figurelabel}
   \end{figure}  

\subsection{Transfer Learning for Domain Adaptation}
Figure 6 presents the SHAP values of an edited image containing a human arm superimposed with a \textit{humanoid} body and which was correctly classified by the model. The image came from VDS2 which was curated with the insights from the well-performing model. This would seem to point towards improved accuracies as the \textit{humanoid} design resembled the human’s more and more since the human hands’ features were used to train the model. But photo-realistic designs are computationally expensive to create. This is an especially relevant concern if applications require very specific gestures or if the gesture classes are to be expanded in the future. Transfer learning may be an effective method. While deep learning requires large datasets, due to the curation of the VEnv dataset leveraging learned features from the REnv, small datasets would be required to achieve high accuracy. Indeed, with just 20 images/class, transfer learning on \textit{model0\_1} yielded classification accuracy of 0.74 on VDS2 test set thus, successfully adapting the model trained on data from the real environment to the virtual one.

\section{Conclusions: Guidelines for Adaptation}
The interpretation of the features learned and expected by the well-performing model dictated the curation of the virtual environment (VEnv) dataset. If transfer learning is determined to be less expensive than adding greater detail to the dataset, the dataset may be expanded to allow for this step. Section xxx demonstrates that only a limited number of images are required for this purpose since the curated dataset is as close to being in-distribution of the original training data as is economically feasible. Some of the features that may be generalizable across datasets containing humans may be the addition of a\textit{ humanoid} with material colour resembling the human skin tone. Apart from that, light sources must be placed in the VEnv so as to mimic the original lighting of the images since computer vision models are notoriously prone to misclassification if lighting conditions differ. Finally, poses and expressions must be designed in an anatomically correct way. In order to make the final model flexible to accommodate the dynamic environment, the\textit{ humanoid} was placed in different parts of the \textit{scene} (centre, left and right offset, close to and further away from the camera)(refer Figure 7 for examples of the final dataset). In conclusion, these generic guidelines can provide a research and design guide for development of VR applications and leverage its possibilities for human-robot collaboration.

   \begin{figure} 
   \centering
      \includegraphics[scale=0.92]{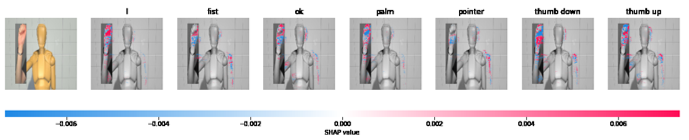}
      \caption{SHAP values on image of \textit{humanoid} with human hand superimposed  which was correctly classified indicating better results with greater photorealism in virtual dataset}
      \label{figurelabel}
   \end{figure}

   \begin{figure} 
   \centering
      \includegraphics[scale=0.6]{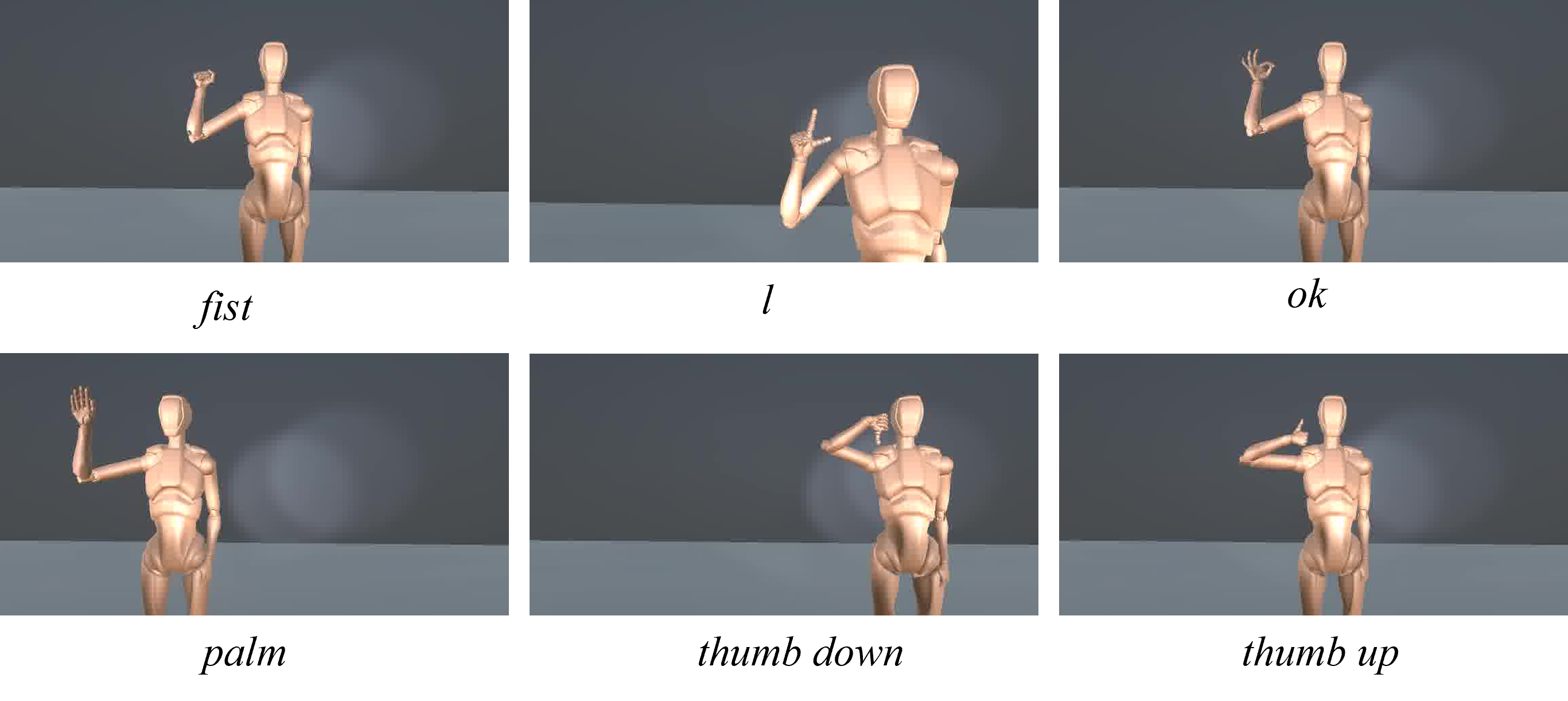}
      \caption{Example images from final VEnv dataset used for transfer learning}
      \label{figurelabel}
   \end{figure}
   
\newpage
\section{Acknowledgements}
Research supported by UBC Office of the Vice-President, Research and Innovation in the form of seed funding to establish research on digitalization of manufacturing and Mitacs Globalink Research Internship Program, 2022.

\bibliographystyle{unsrtnat}
\bibliography{references}  






\end{document}